\documentclass[prl,twocolumn,superscriptaddress]{revtex4}%
\usepackage{amsfonts}
\usepackage{amsmath}
\usepackage{amssymb}
\usepackage{graphicx}%
\begin{document}
\preprint{cond-mat/?????}
\title{Campbell Penetration Depth of a Superconductor in the Critical State}

\author{R.~Prozorov}
\affiliation{Loomis Laboratory of Physics, University of Illinois at Urbana - Champaign, 1110 West Green Street, Urbana, Illinois 61801.}
\affiliation{Department of Physics \& Astronomy and USC NanoCenter, University of South Carolina, 712 Main St, Columbia, SC 29208.}

\author{R.~W.~Giannetta}
\affiliation{Loomis Laboratory of Physics, University of Illinois at Urbana - Champaign, 1110 West Green Street, Urbana, Illinois 61801.}

\author{N.~Kameda}
\affiliation{Department of Applied Physics, The University of Tokyo, Hongo, Bunkyo-ku, Tokyo 113-8656, Japan.}

\author{T.~Tamegai}
\affiliation{Department of Applied Physics, The University of Tokyo, Hongo, Bunkyo-ku, Tokyo 113-8656, Japan.}

\author{J.~A.~Schlueter}
\affiliation{Materials Science Division, Argonne National Laboratory, Argonne, Illinois 60439.}

\author{P.~Fournier}
\affiliation{D\'{e}partement de Physique Universit\'{e} de Sherbrooke Sherbrooke, Quebec, Canada, J1K 2R1.}



\begin{abstract}
The magnetic penetration depth $\lambda(T,H,j)$ was measured in the presence of a slowly relaxing supercurrent, $j$. In single crystal
$\mathrm{Bi_2Sr_2CaCu_2O_8}$ below approximately 25 K, $\lambda(T,H,j)$ is strongly hysteretic. We propose that the irreversibility arises from a
shift of the vortex position within its pinning well as $j$ changes. The Campbell length depends upon the ratio $j/j_{c}$ where $j_{c}$ is the
critical current defined through the Labusch parameter.  Similar effects were observed in other cuprates and in an organic superconductor.

\end{abstract}
\volumeyear{year} \volumenumber{number} \issuenumber{number}
\date{November 2002}
%\received[Received text]{date}%
%\revised[Revised text]{date}%
%\accepted[Accepted text]{date}%
%\published[Published text]{date}

\maketitle

Many measurements have shown that the character of vortex pinning in BSCCO changes qualitatively in region of $20-30$ K
\cite{niderost,khaykovich,goffman,correa2001,rodriguez,dewhurst,tamegai}. The second magnetization peak disappears and the critical current increases
sharply.  The Larkin pinning length \cite{larkin1979} becomes comparable to the interplanar spacing implying that vortex pancakes are pinned
individually ($0-D$ pinning) rather than as components of an elastic string. In this region the ac susceptibility measured in a zero-field cooled
(ZFC) sample differs markedly from that obtained in a field cooled (FC) sample \cite{rodriguez}.  ZFC samples represent a nonequilibrium flux profile
and the small signal response of such a system is not fully understood \cite{geshkenbein,gurevich}. In this paper we report measurements of the
penetration depth in both FC and ZFC samples. Our measurements show strong hysteresis and memory effects but are not in the limit of strong driving
fields where the ac field itself can induce new vortex phases \cite{gordeev}. We propose that the hysteretic ZFC response can be understood as a
generalized Campbell penetration depth $\lambda_C(B,T,j)$ that depends upon the slowly relaxing supercurrent $j$ as well as the curvature of the
pinning potential as parameterized by $j_c$.  We compare data in $\mathrm{Bi_2Sr_2CaCu_2O_8}$ (BSCCO, $T_c\sim 92$ K) to measurements in
electron-doped $\mathrm{Pr_{1.85}Ce_{0.15}CuO_4}$ (PCCO, $T_c\sim 24$ K), an organic superconductor
$\mathrm{\beta^{\prime\prime}-(ET)_2SF_5CH_2CF_2SO_3}$ ($\beta^{\prime\prime}$-ET, $T_c\sim 5$ K) \cite{geiser} and $Nb$ ($T_c\sim 9.3$ K).
\begin{figure}[tb]
\includegraphics[width=8cm]{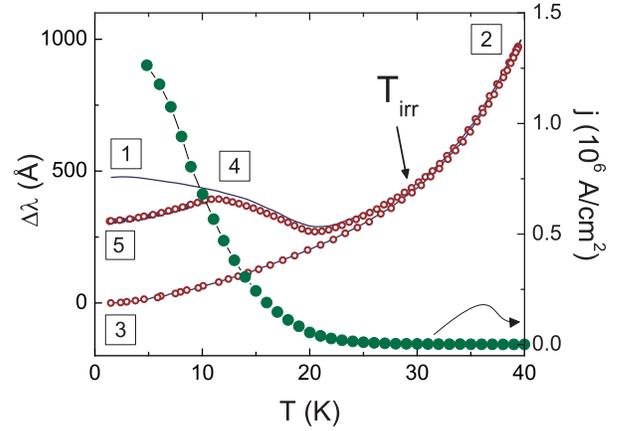}%
\caption{\underline{Experiment $1$:} BSCCO crystal ZFC to $1.5$ K and $H_{dc}=260$ Oe applied.  Signal followed curve $1\rightarrow2\rightarrow3$.
\underline{Experiment $2$:} BSCCO crystal ZFC to $12$ K and $H_{dc}=260$ Oe applied.  Sample was then cooled and warmed following the path
$4\rightarrow5\rightarrow4\rightarrow2\rightarrow3$.  Right axis shows current density estimated from the irreversible magnetization.}%
\label{lac}%
\end{figure}
The penetration depth was measured with an $11\ MHz$ tunnel-diode driven LC resonator \cite{resonator,prozorov} mounted in a $^{3}He$ refrigerator.
An external dc magnetic field ($0-7$ kOe) was applied parallel to the ac field ($\sim5$ mOe). The oscillator frequency shift $\Delta
f=f(T)-f(T_{min})$ is proportional to the ac susceptibility and, therefore, to the change in penetration depth,
$\Delta\lambda=\lambda(B,T)-\lambda(B,T_{min})$ via $\Delta f=-G\Delta\lambda$, where $G$ is a calibration constant \cite{resonator,prozorov}. For an
ac magnetic field along the c-axis, only ab-plane rf screening currents are excited. Although this results in a much smaller ac Lorentz force on
vortices, it removes complications from interplane currents.  The absence of c-axis currents is demonstrated by the zero field data in BSCCO. We
obtain a linear change $d\lambda_L/dT\approx11$ \AA/K in good agreement with earlier measurements and indicative of a $d-$wave superconductor
\cite{jacobs,shibauchi,prozorov2}. Any significant tilt of the ac field would generate c-axis currents and give a much larger value of
$d\lambda_L/dT$.
The total penetration depth in the mixed state is $\lambda^2=\lambda_L^2+\lambda_{vortex}^2$ where $\lambda_L$ is the London penetration depth and
$\lambda_{vortex}$ the contribution from vortex motion.  A comprehensive expression for $\lambda_{vortex}$ has been derived by several authors
\cite{clem,brandtac,beek}.  At low temperatures and frequencies well below the pinning frequency (of order GHz in cuprates), $\lambda_{vortex}$
reduces to the Campbell pinning penetration depth $\lambda_C^2=\phi_0 B/4\pi\alpha$ \cite{campbell}. Here $\phi_0$ is the flux quantum and $\alpha$
is the Labusch parameter \cite{labusch}.  Our measurements give $\alpha > 10^3$ dyne/cm$^2$ for temperatures below $25$ K, so the maximum vortex
excursion due to the ac current is less than $5$ \AA.  This value is well within the range of individual pinning wells (of order a coherence length
or more), justifying our assumption of small oscillations.
\begin{figure}[ptb]
\includegraphics[width=8cm]{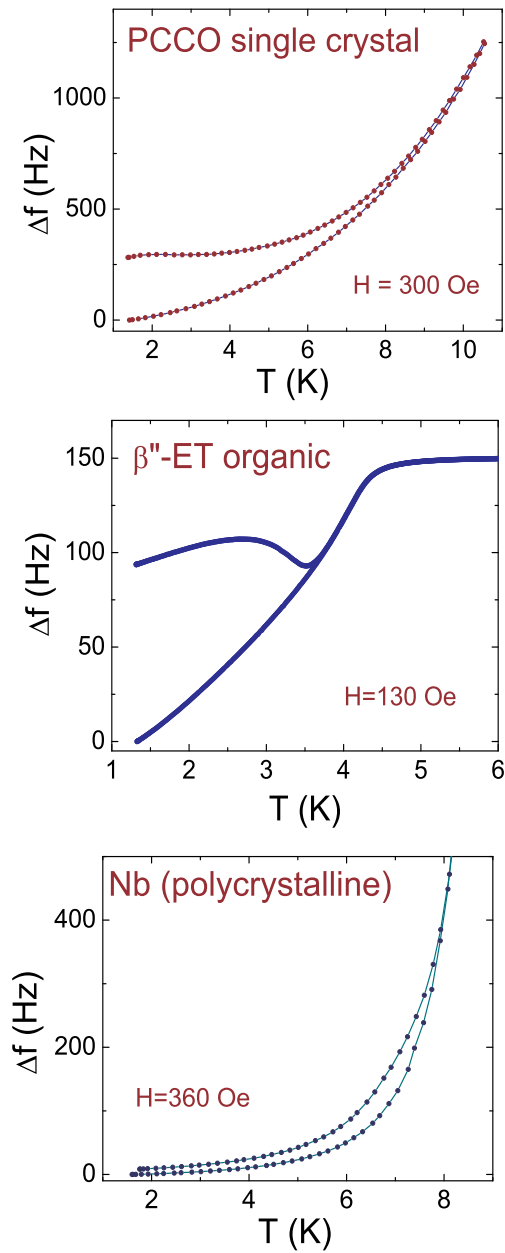}%
\caption{FC and ZFC frequency shift (proportional to penetration depth) in  single crystals of PCCO, $\beta^{\prime\prime}$-ET, and polycrystalline
$Nb$.}
\label{panel}%
\end{figure}
Fig.~\ref{lac} presents $\lambda(B,T)$ for a BSCCO single crystal as the temperature was cycled. After zero field cooling (ZFC) to $1.5$ K, the dc
field was ramped from $0\rightarrow-7\ kOe\rightarrow 260$ Oe. This procedure ensured that the entire sample was filled with vortices, but
nonuniformly. The sample was then warmed ($1\rightarrow2$) during which $\lambda$ first decreased then increased again. (Neglecting the initial ramp
to $-7$ kOe yields nearly the same curve but with a weak maximum near $2$ K which may come from portions of the sample where no vortices exist and
thus screen more efficiently). During this phase of the cycle $j$ relaxes as the flux distribution becomes more uniform. On the same plot we show the
screening current $j$ measured on the same sample, in the same field in a SQUID magnetometer. $j$ was determined from the irreversible component of
magnetization and applying the Bean model. $j$ measured in this way is considerably different from $j_{c}$ owing to strong flux creep in the
cuprates. Once the temperature exceeds $T_{irr} \sim 25-30$ K, $j$ relaxes more rapidly and the flux profile becomes uniform. Subsequent cooling and
warming traces ($2\rightarrow3\rightarrow2$) were perfectly reversible and represent the penetration depth of a uniform flux profile. This reversible
curve was identical to that obtained in a field-cooled (FC) experiment and we refer to them interchangeably. The hysteresis between points $1$ and
$3$ in Fig.~\ref{lac} corresponds to a change in rf magnetization of $\leq10^{-7}$ emu, which is at the detectability limit of commercial
magnetometers.

\begin{figure}[ptb]
\includegraphics[width=8cm]{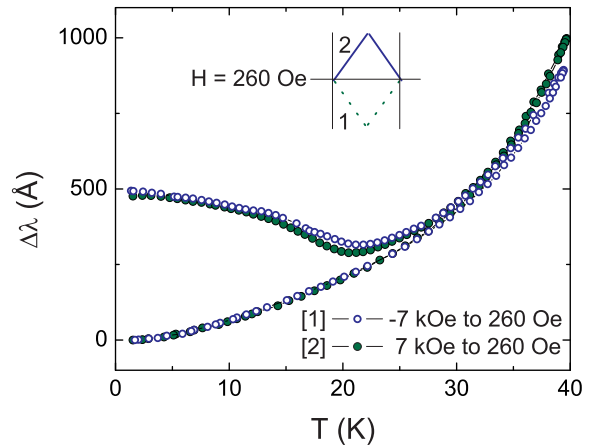}%
\caption{Comparison of $\Delta\lambda(T)$ for flux entry and exit. \textbf{Closed symbols:} magnetic field was ramped up from $-7$ kOe to $+260$ Oe
(flux entry). \textbf{Open symbols:} field was ramped down from $+7$ kOe to $+260$ Oe (flux exit) and the sample was warmed-then-cooled. Schematics
shows the corresponding profiles of vortex density.}%
\label{updn}%
\end{figure}

In Fig.~\ref{panel} we show similar measurements in the electron-doped cuprate PCCO ($\gamma=30-80$), an organic superconductor
$\beta^{\prime\prime}$-ET ($\gamma=400-800$) \cite{prozorov3}, and polycrystalline $Nb$ ($\gamma=1$). Together with BSCCO ($\gamma=300-400$), these
materials span a wide range in transition temperature and anisotropy $\gamma$.  All three anisotropic superconductors show nonmonotonic ZFC
temperature dependence, represented by the top curve in each panel. By contrast, in Nb and in YBCO (also measured but not shown, $\gamma=6-8$)
$\lambda(ZFC)$ always increases monotonically with temperature. Returning to Fig.~\ref{lac}, although the pinning changes dramatically near $25$ K,
the change in the penetration depth is observable only in the ZFC curve. The FC curve is perfectly smooth. Goffman \textit{et al.} \cite{goffman}
reported measurements of the transverse susceptibility (ac field along the ab - plane) at very low frequencies that \textit{did} show a sharp
increase in screening below $22$ K.  This feature disappeared at kHz frequencies \cite{goffman}, which is consistent with our data at $11$ MHz. We
now focus on the ZFC behavior.
\begin{figure}[ptb]
\includegraphics[width=8cm]{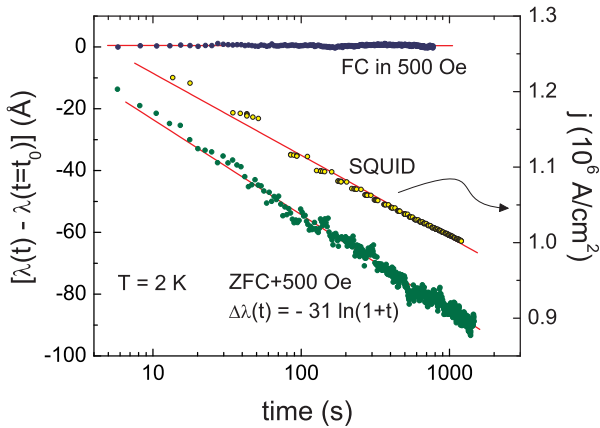}%
\caption{Time-logarithmic relaxation of the ac penetration depth after application of a $500$ Oe magnetic field at $2$ K (lower curve) and after
FC in $500$ Oe (upper curve). Right axis refers to the relaxation of the current $j$ obtained from the magnetization measurements.}%
\label{creep}%
\end{figure}
Fig.~\ref{creep} shows the time dependence of various quantities. The topmost curve shows $\lambda$ when the sample field was cooled to $2$ K in
$H_{dc}=500$ Oe. Relaxation is negligible. When the sample was zero-field cooled and then $500$ Oe applied, $\lambda$ and $j$ showed logarithmic
relaxation.  This correspondence suggests that in a ZFC state, the penetration depth should have a direct functional relationship to $j$. This is
confirmed in Fig.~\ref{updn} where we compare $\lambda$ measured at the same final value of field but with two entirely different flux profiles. The
solid symbols correspond to the initial application (at $1.5$ K) of a $-7$ kOe magnetizing field, as before, while the open symbols correspond to a
$+7$ kOe magnetizing field. Both fields were then returned to $H=+260$ Oe before the temperature sweep began. The distribution of $B$ throughout the
sample was entirely different for these two starting conditions, as shown schematically. However, within a critical state picture, the magnitude of
$dB/dx$ and thus $j$ remains the same for these two distributions. The fact that $\lambda$ vs. $T$ was unchanged for the two starting conditions is
strong evidence that $j$ and not $B$ is the determinant of the penetration depth in the nonuniform state.  Some implicit dependence of the Labusch
constant upon $B$ could account for minor differences between the curves at higher temperatures.
Based on these results, we propose the following model for $\lambda_C(B,T,j)$ in a superconductor with a non-uniform flux profile.  The supercurrent
$j$ biases vortices away from equilibrium through the Lorentz force, $F_L=j\times \phi_0/c$. The Campbell depth is then determined by the curvature
of the pinning potential well at the biased position. For a pinning potential, $V\left(  r\right)  $, the vortex displacement, $r_{0}$, is found from
$dV/dr=F_{L}$.  The maximum force determines the critical current, $j_c=c\alpha_0r_p/{\phi_0}$, attained at the range of the pinning potential
$r_{p}$. The effective Labusch constant $\alpha\left(j\right)$ is then determined from $\alpha\left(j\right)=\left. d^2V/dr^2\right\vert_{r=r_{0}}.$
For example, consider the form $V(x)=\alpha_{0}(T)x^{2}\left(1-x/3\right)/2$, for which the volume pinning force saturates. Here $x=r/r_{p}$ is a
dimensionless vortex displacement. This potential has been used to analyze the quantum tunneling of vortices \cite{caldiera}. The supercurrent $j$
biases the vortex segment to a new position $x_{0}=1-\sqrt{1-j/j_{c}}$ where the local curvature is $\alpha(j)=\alpha_{0}\sqrt{1-j/j_{c}}$. The
change in curvature produces a $j$ dependence to the Campbell depth:
\begin{equation}
\lambda_C^2=\frac{\phi_{0} B}{4\pi\alpha(j)}=\frac{ \phi_{0} B}{4\pi\alpha_{0}} \frac{1}{\sqrt{1-j/j_c}}=\frac{\lambda_C^2\left( j = 0\right) }
{\sqrt{1-j/j_{c}}} \label{lcamp}%
\end{equation}
The model predicts that $\lambda_{C}(ZFC) > \lambda_{C}(FC)$ since $j = 0$ in the FC case.  This conclusion remains true for other pinning potentials
such as $V\left(  x\right) =\alpha_{0}x^{2}\left(  1-x^{2}/6\right)/2$ \cite{yamajuji}.  As Fig.~\ref{panel} shows, $\lambda_C(ZFC) > \lambda_C(FC)$
in all materials studied. The model also predicts that $j/j_c$, and not $B$ explicitly, determines the nonequilibrium component of the penetration
depth, as shown in Fig.~\ref{updn}.
Rodriguez \textit{et al.} \cite{rodriguez} have previously reported a difference in the ac susceptiblity between FC and ZFC samples of BSCCO. Their
data look similar to ours, although they observed hysteresis only when the ac field was \textit{parallel} to the conducting planes, inducing both
ab-plane and c-axis currents.  They also worked at considerably larger ac field amplitudes.  They attributed the nonmonotonic ZFC curve to a c-axis
critical current that was non-monotonic with temperature in the critical state. This effect presumably occurs only in highly anisotropic materials
like BSCCO.  We found nonmonotonic ZFC curves only in the highly anisotropic materials studied (BSCCO, PCCO and $\beta^{\prime\prime}$-ET) so two
dimensionality clearly is important as those authors emphasize.  However, our model involves no c-axis currents and shows that the current biasing
effect predicts $\lambda_{C}(ZFC) > \lambda_{C}(FC)$ as observed. The precise shape of the ZFC curve depends upon how rapidly $j$ relaxes during the
sweep and the thermal history of the sample, so it is not a basic property of the superconductor.  For example, in Fig.~\ref{lac}, if we ZFC only to
$12$ K instead of $1.5$ K, the maximum in $\lambda$ occurs at point $4$, i.e., the lowest temperature achieved on the initial cooldown. We then trace
the path $4\rightarrow 5\rightarrow 4\rightarrow 2\rightarrow 3$.  Despite being in a nonequilibrium state, the system retains perfect memory in its
passage from point $4$ to $5$ and back to $4$.

\begin{figure}[ptb]
\includegraphics[width=8cm]{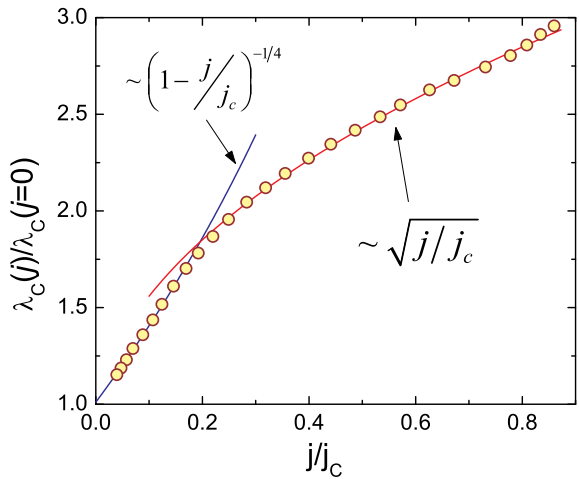}
\caption{Experimentally determined $\lambda_{C}\left(j/j_{c}\right)/\lambda_C(0)$ for
BSCCO single crystal}%
\label{lvsj}%
\end{figure}

In principle the $j/j_c$ dependence of $\lambda_C(B,T,j)$ allows one to determine the shape of the pinning potential. With $\lambda_L \left( 0
\right)\approx 2690$ \AA, determined from our earlier measurements \cite{prozorov2} we obtain total penetration depth $\lambda_C(B,T,j)$.
Fig.~\ref{lvsj} shows $\lambda_C (j/j_c)$ for BSCCO taken from data points at different temperatures and thus different values of $j/j_{c}$. $j_c$ is
obtained from force balance, $4\pi j_c/c=Br_p/\lambda_C^2\left(0\right)$ with $r_p = 30$ \AA~ where $\lambda_C(j = 0)$ is obtained from the FC data.
Up to $j/j_c = 0.2$ the data fits Eq.(1) rather well, but at higher values of $j/j_c$ the dependence is much weaker, closer to $\sqrt{j/j_c}$.  The
larger values of $j/j_c$ correspond to lower temperature points, deep inside the $0D$ pinning region.  There is no explicit form for the pinning
potential is this glassy phase and in fact this experiment may be the first attempt to measure it. Interestingly, the observed $\sqrt{j/j_c}$
behavior follows from a nonanalytic pinning potential $V(x) \propto x^{3/2}$ but we have no theoretical justification for this form.
In conclusion, we propose a current biasing effect to explain the difference between FC and ZFC measurements of the Campbell penetration depth in a
variety of superconductors.  The supercurrent $j$ biases vortices to a new position in the pinning potential and the Campbell depth measures the
local curvature, which depends upon $j/j_c$.  This effect may be used to probe the shape of the vortex pinning potential.

Acknowledgements: We thank V.~Geshkenbein, A.~Koshelev, V.~Vinokur, and J.~R.~Clem for useful discussions. Work at UIUC was supported by NSF DMR
0101872. Work at USC was supported by the NSF/EPSCoR under Grant No. EPS-0296165. Work at UT The Grant-in-Aid for Scientific Research from the
Ministry of Education, Culture, Sports, Science, and Technology. Research at Argonne was supported by DOE, Office of Basic Energy Sciences, Division
of Materials Sciences, under contract No. W-31-109-ENG-38.

\end{document}